\documentstyle[sprocl,epsfig,wrapfig]{article}

\def\be{\begin{equation}}
\def\ee{\end{equation}}
\def\bea{\begin{eqnarray}}
\def\eea{\end{eqnarray}}

\def\EEJET{E^{\rm j}_{\rm e}}
\def\EJET{E^{\rm j}}
\def\Evis{E_{\rm vis}}
\def\njet{n_{\rm j}}
\def\Mjj{M_{\rm jj}}
\def\etajet{\eta^{\rm j}}

\def\Ee{E_{\rm e}}

\def\ptjet{E_{\rm T}^{\rm j}}

\def\ee{\mbox{e}^+\mbox{e}^-}
\def\WW{\mbox{W}^+\mbox{W}^-}

\def\Z0{\mbox{Z}^0}
\def\LQ{\mbox{LQ}}
\def\X{\mbox{X}}
\def\ETMISS{E_{\rm T}\!\!\!\!\!\!\!/\;\;}

\def\qmax{Q^2_{\rm max}}

\def\qqbar{\mbox{q}\overline{\mbox{q}}}
\def\fg{f_{\gamma/e}}

\begin{document}

\title{SEARCH FOR SINGLE LEPTOQUARK PRODUCTION \\ IN
ELECTRON-PHOTON SCATTERING \\ AT \boldmath $\sqrt{s}=161$ 
AND $172$~GEV \footnotemark[1] 
\unboldmath}


\author{ STEFAN S\"OLDNER-REMBOLD \\ for the OPAL collaboration}

\address{Universit\"at Freiburg, Hermann-Herder-Str.~3,\\
D-79104 Freiburg i.~Br., Germany}


\maketitle\abstracts{
A search for a first generation scalar leptoquark (LQ) has been performed using
the data collected by the OPAL detector
in 1996 at $\ee$ centre-of-mass energies $\sqrt{s}$
of 161 and 172~GeV. It is assumed that a single leptoquark
can be produced in the process eq$\rightarrow\LQ$, where 
the initial state quark originates from a hadronic fluctuation of a
quasi-real photon which has been radiated by one of the LEP beams.
Lower limits at the 95~\% confidence level 
on the mass of a first generation scalar leptoquark of
$131$~GeV for $\beta=0.5$ and $\beta=1$, coupling values $\lambda$ 
larger than $\sqrt{4\pi\alpha_{\rm em}}$ and leptoquark charges
$-1/3$ or $-5/3$ are obtained.}

\section{Kinematics and Monte Carlo simulations}
\footnotetext[1]{To be published in the proceedings of PHOTON'97, Egmond aan Zee}
Leptoquarks are coloured spin 0 or spin 1 particles carrying both
baryon and lepton quantum numbers. 
Recently it has been suggested to search
for leptoquarks in electron-photon collisions at LEP \cite{bib-doncheski}.
The photon, which has been radiated by one of the LEP beams, serves
as a source of quarks through its fluctuations into hadronic states.
The electron-quark interaction produces a leptoquark which is 
assumed to decay subsequently into an electron or a neutrino 
and a quark.\footnotemark[2] 
\footnotetext[2]{Charge conjugation is
implied throughout this paper and positrons are referred to as electrons}

In electron-photon scattering first generation leptoquarks of
charge $-1/3$, $-5/3$, $-2/3$ and $-4/3$ can be produced.
The cross-section to produce charge $-2/3$ and $-4/3$ leptoquarks
is suppressed, since there is less d quark content in the photon than
u quark. The limits will therefore be given for 
leptoquark charges $-1/3$ or $-5/3$.
The cross-sections in e$\gamma$ scattering for both charge states 
are identical, since it is equally probable to find a u or a
$\overline{\mbox{u}}$ quark in the photon. In principle this search
is also sensitive to electron-charm states, since the probability
for a photon to split into c$\overline{\mbox{c}}$ or
u$\overline{\mbox{u}}$ is
expected to be about equal for leptoquark masses $M>>m_{\rm c}$.
Furthermore it has been assumed that either left or
right handed couplings to fermions vanish.
The cross-sections in e$\gamma$ scattering for both couplings
are identical, whereas the branching ratio $\beta$ into eq final states is 1 for
right handed couplings and 1/2 for left handed couplings~\cite{bib-buch}.

The total cross-section for the production of scalar leptoquarks of mass $M$ is
a convolution 
of the Weizs\"acker-Williams effective photon
distribution $\fg(z)$, with $z$ being the momentum fraction carried
by the photon,
and the parton distribution functions $f_{q/\gamma}(x,\mu^2)$ 
of the photon, evaluated at
the scale $\mu=M$ \cite{bib-doncheski}:
\begin{equation}
\sigma(\ee\rightarrow\LQ + \X) =\frac{\lambda^2\pi}{2s}
\int_{M^2/s}^1\frac{dz}{z}f_{\gamma/e}(z,\qmax)
f_{q/\gamma}(M^2/(zs),M^2).
\end{equation}
The Monte Carlo simulation of this process
is done with PYTHIA 5.722 \cite{bib-pythia,bib-lqpythia}.
In the simulation the maximum photon virtuality $\qmax$ 
used in the Equivalent Photon Approximation equals $s/4$,
but the simulated photon is always real ($Q^2=0$).
The GRV parametrisation \cite{bib-grv} of the parton distribution functions
was used. In the kinematic region relevant for leptoquark
production the variations of the cross-section due to the different
parameterisations are small.
Interference effects with deep-inelastic e$\gamma$ scattering
are also neglected. The total cross-section in
PYTHIA for $\sqrt{s}=172$ is about 10--20~\% lower than the cross-sections  
given for $\sqrt{s}=175$ in Ref.~1.
Vector leptoquarks can currently not be simulated with PYTHIA.
The limits are therefore given only for scalar 
leptoquarks. The standard PYTHIA Monte Carlo has been modified
to include $\LQ\rightarrow\nu_{\rm e}$d  decays in addition to the standard 
$\LQ\rightarrow$~eu decays.
\section{Event Analysis}
\label{sec-evsel}
Jets were reconstructed using
a cone jet finding algorithm with a cone size $R=1$ and 
a cut on the minimum transverse jet energy $E_{\rm T}$ of
15 GeV. 
Tracks and calorimeter clusters 
were used as input for the jet finding algorithm and for determining
the missing transverse energy $\ETMISS$ of the event. A matching algorithm
between tracks and clusters is applied. 
The electron was identified using the standard OPAL
neural net electron identification~\cite{bib-elid}. 
All relevant Standard Model background processes were studied using 
Monte Carlo generators. 
The total data sample corresponds to an integrated luminosity of
20.5 pb$^{-1}$.

\subsection{The electron plus hadronic jet channel}
For this channel the identified electron with
the largest momentum was assumed to be the electron from the leptoquark
decay. The electron is
usually reconstructed as a jet.
Candidate events were selected based on
the following cuts:
\begin{itemize}
\item In order to reduce background from deep-inelastic
e$\gamma$ and $\ee\rightarrow\tau^+\tau^-$ events,
exactly two jets must have been found in the event ($\njet=2$).
\item A large number of $\ee\rightarrow\tau^+\tau^-$ and
$\ee\rightarrow\ee$ events are rejected by requiring
a minimum number of 5 reconstructed tracks ($n_{\rm ch} \ge 5$).
In addition, the ratio $E_{\rm ECAL}/\sqrt{s}$ has to be less than 0.9,
where $E_{\rm ECAL}$ is the energy in the electromagnetic calorimeter.
\item 
The missing transverse energy $\ETMISS$  must be less than 15 GeV in order
to reduce background from $\tau^+\tau^-$ and $\WW$ pair production. 
\item 
An isolation cut is applied on the identified electron. 
The jet with the smallest angular distance to the electron
is chosen to be the electron jet. The
difference between the energy $\EEJET$ of this jet and
the energy $\Ee$ of the electron must be less than 4~GeV.
Most multihadronic $\ee\rightarrow\qqbar$
events are removed by this cut. 
\item Events where an electron was scattered at a small
angle are
rejected by requiring for the angle of the electron 
$|\cos \theta_{\rm e}| <0.85$.
\item 
The total multiplicity $n_{\rm q}$ of the quark 
jet must be $n_{\rm q}\ge 7$, where $n_{\rm q}$
is the total number of tracks  
and calorimeter clusters associated to this jet.
\end{itemize}
The cuts on the transverse momenta of
the jets and on the angle $|\cos \theta_{\rm e} |$ of
the electron reduce significantly the sensitivity to find 
a leptoquark which is lighter than approximately $M_{\rm Z}/2$, 
the region excluded by the LEP1 searches.
These cuts are necessary to reduce
the background from deep-inelastic e$\gamma$ 
events which becomes increasingly important at small masses.

After all cuts we expect a background 
of $5.2\pm0.4$ events from all Standard Model processes. 
In the data four events are observed with jet-jet invariant masses 
$\Mjj$ of 36, 37, 62 and 98~GeV. 
In Fig.~\ref{fig-mnq}a the 
$\Mjj$ distribution of the four candidate events is shown
together with the sum of all Monte Carlo background distributions. 
Also shown is a possible leptoquark signal for 
$\lambda=\sqrt{4\pi\alpha_{\rm em}}$ and different LQ masses. 
The mass distribution of the candidate events is 
consistent with the expectation from the background Monte Carlo simulation.

\subsection{The neutrino plus hadronic jet channel}
This search has to be
optimized for a single hadronic jet in the detector. Its
transverse energy $\ptjet$ must be balanced by the neutrino.
The cuts are therefore:
\begin{itemize}
\item In order to reject events with large missing transverse
      energy due to badly measured tracks, the ratio
      of the energy $E_{\rm ECAL}$ to the total visible
      energy $\Evis$ has to be larger than 20~\%.
\item Exactly one jet has to be found ($\njet=1$) with
      $\ptjet>15$~GeV.
      The jet direction in the laboratory frame
      is required to lie within a pseudorapidity range $|\etajet|<2$
      to reject events where a single jet, usually
      due to an electron, was found in the forward
      detectors.
\item $n_{\rm ch}\ge 5$ and $n_{\rm q}\ge 7$.
\item The missing transverse energy $\ETMISS$ must be greater than 15~GeV
      and it should be mainly due to the jet.
      Therefore we require $|\ptjet-\ETMISS|<3$ GeV
      and $\EJET/\Evis>0.5$.
\end{itemize}
Since no additional cuts on electron variables
are necessary, the efficiency
to detect a leptoquark is higher in the $\nu_{\rm e}$q$'$ than in
the eq channel. For $M=100$~GeV the efficiency is about 61~\% in
the $\nu_{\rm e}$q$'$ and 55~\% in the eq channel.

\begin{figure}[htbp]
\begin{tabular}{ccc}
\epsfig{file=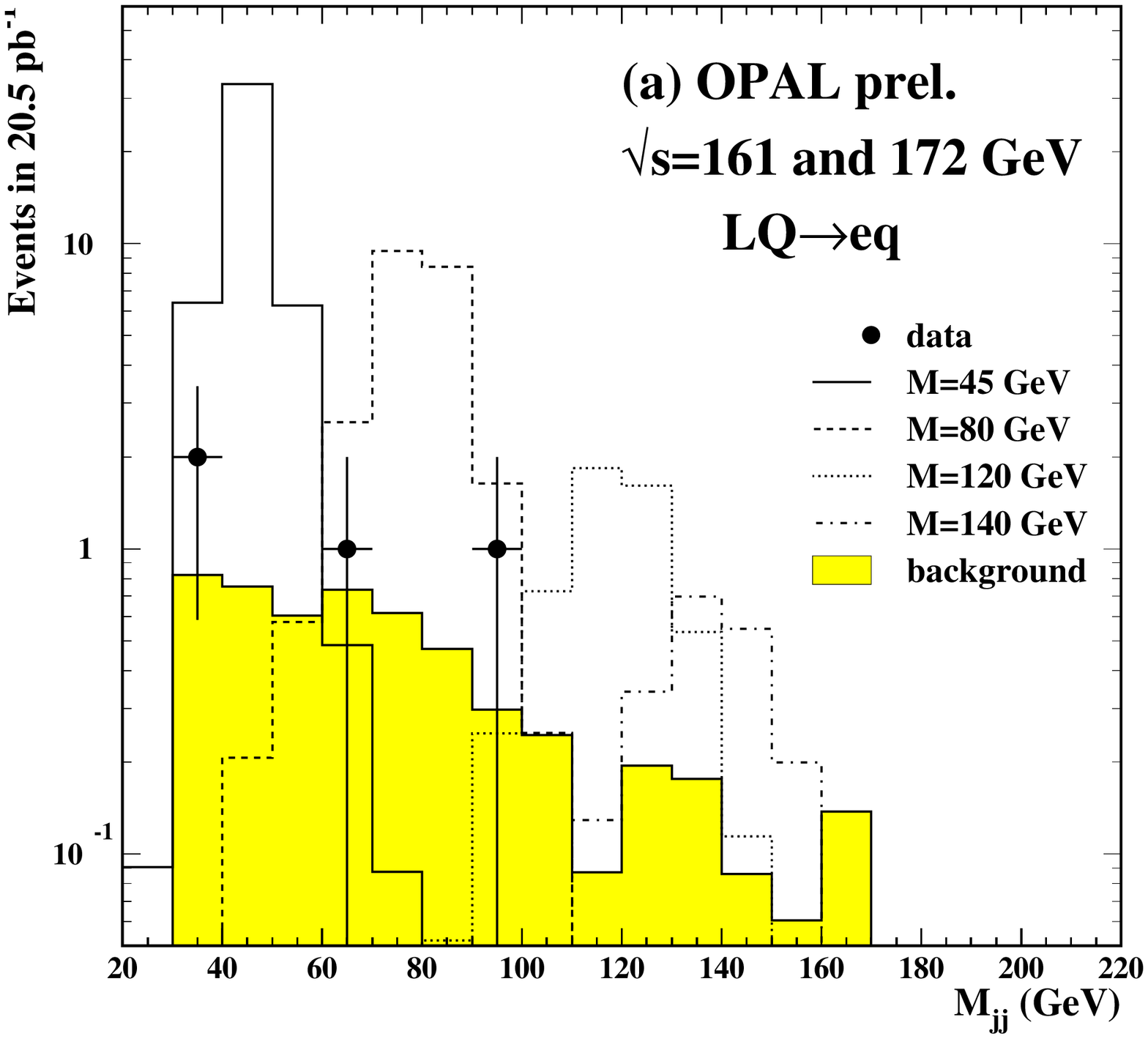,
width=0.46\textwidth,height=5.5cm} &
\epsfig{file=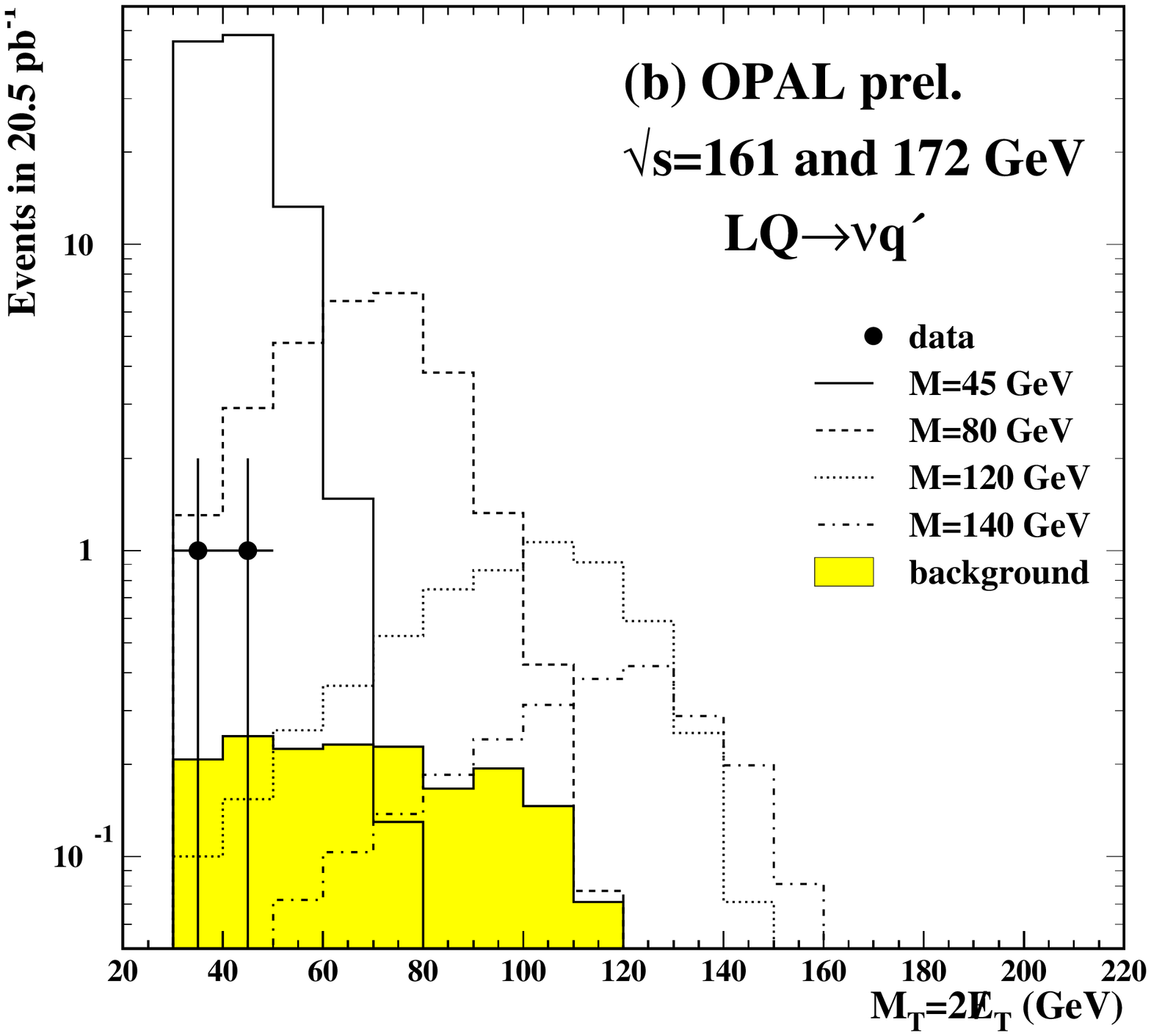,
width=0.46\textwidth,height=5.5cm} &
\end{tabular}
\caption{Number of (a) $\LQ\rightarrow$~eq and
(b) $\LQ\rightarrow\nu_{\rm e}$q$'$ events expected 
with $\lambda=\sqrt{4\pi\alpha_{\rm em}}$ in 20.5~pb$^{-1}$ of data
after all cuts for $M=45, 80, 120$ and $140$~GeV
(histograms) and the candidate events (data points). 
The sum of all background contributions expected from the simulation
of the Standard Model processes is also shown normalized to the data
luminosity.} 
\label{fig-mnq}
\end{figure}
The leptoquark mass was reconstructed by calculating the 
transverse mass $M_{\rm T}=2\ETMISS$.
The transverse mass $M_{\rm T}$ of the two candidate events at 38 and 46 GeV
is shown in Fig.~\ref{fig-mnq}b
together with the background distribution from the simulation.
The expected background rate is $1.81\pm0.05$ events.
The transverse mass distribution for a leptoquark
production cross-section using $\lambda=\sqrt{4\pi\alpha_{\rm em}}$ 
is also indicated.
\section{Mass limit for a scalar leptoquark}
\begin{wrapfigure}{r}{6.6cm}
\epsfig{figure=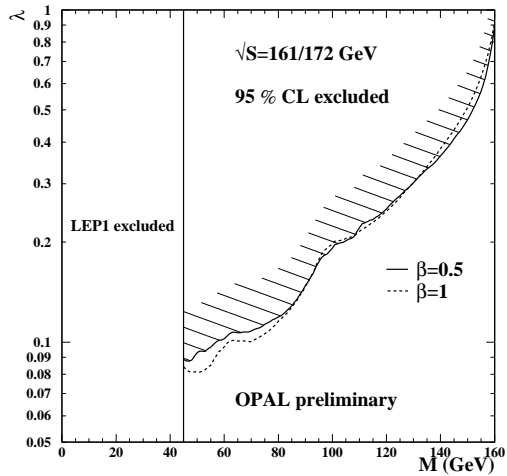,width=6.6cm}
\label{fig-limit}
\caption{Upper limit at 95~\% CL of the coupling
$\lambda$ as a function of the mass $M$ of the scalar LQ.}
\end{wrapfigure}
The systematic error includes
(a) the luminosity measurement with 1~\%, (b)
the model dependence of the leptoquark fragmentation  with
4~\%, (c) the electron identification efficiency with 2~\% and
(d) the Monte Carlo statistics with 1~\%. The model dependence
of the leptoquark fragmentation was estimated by varying
the cut on the average charged multiplicity by one unit in the
Monte Carlo while keeping it fixed in the data.

The limit was obtained separately for $\beta=1$ and for $\beta=0.5$.
The 95~\% confidence level (CL) upper limit was calculated taking
into account the candidates, the background, the experimental
resolution and the systematic errors. The cross-section
was determined using PYTHIA.
The upper limit at 95~\% CL of the coupling
$\lambda$ as a function of the leptoquark mass $M$ is
given in Fig.~2. The mass limits are 
$M>131$~GeV for both $\beta=0.5$ and $\beta=1$ and
for $\lambda=\sqrt{4\pi\alpha_{\rm em}}$.

\end{document}